\documentclass[aps,pre,preprint,groupedaddress,showpacs]{revtex4-1}

\usepackage[pdftex]{graphicx}
\usepackage{amssymb,amsfonts,amsmath,natbib}

\begin{document}

\title{The persistence of social signatures in human communication}

\author{J.~Saram\"aki}
\affiliation{Department of Biomedical Engineering and Computational Science, Aalto University School of Science,
00076 AALTO, Espoo, Finland}
\author{E.A.~Leicht}
\affiliation{CABDyN Complexity Center, Sa\"{i}d Business School, University of Oxford,  OX1 1HP, UK}
\author{E. L\'opez}
\affiliation{CABDyN Complexity Center, Sa\"{i}d Business School, University of Oxford,  OX1 1HP, UK}
\author{S.G.B.~Roberts}
\affiliation{Department of Psychology, University of Chester, CH1 4BJ, UK}
\author{F.~Reed-Tsochas}
\affiliation{CABDyN Complexity Center, Sa\"{i}d Business School, University of Oxford,  OX1 1HP, UK}
\affiliation{Department of Sociology, University of Oxford, Oxford OX1 3UQ, UK}
\author{R.I.M.~Dunbar}
\affiliation{Department of Experimental Psychology, University of Oxford, OX1 3UD, UK}



\begin{abstract}
The social network maintained by a focal individual, or ego, is intrinsically dynamic and typically exhibits some turnover in membership over time as personal circumstances change. However, the consequences of such changes on the distribution of an ego's network ties are not well understood.  Here we use a unique 18-month data set that combines mobile phone calls and survey data to track changes in the ego networks and communication patterns of students making the transition from school to university or work. Our analysis reveals that individuals display a distinctive and robust social signature, captured by how interactions are distributed across different alters. Notably, for a given ego, these social signatures tend to persist over time, despite considerable turnover in the identity of alters in the ego network. Thus as new network members are added, some old network members are either replaced or receive fewer calls, preserving the overall distribution of calls across network members. This is likely to reflect the consequences of finite resources such as the time available for communication, the cognitive and emotional effort required to sustain close relationships, and the ability to make emotional investments.
\end{abstract}

\maketitle

\section{Introduction}

Social relationships play an important functional role in human society both at the collective level and by providing benefits to individuals. In particular, it appears that having strong and supportive relationships, characterized by closeness and emotional intensity, is essential for health and wellbeing in both humans and other primates~\cite{holt2010,wittig2008}. At the same time, there is a higher cost to maintaining closer relationships, reflected in the amount of effort required to maintain a relation at the desired level of emotional closeness. Because of this, the number of emotionally intense relationships is typically small. Moreover, it has been suggested that ego networks, the sets of ties individuals (egos) have to their friends and family (alters), may be subject to more general constraints associated with limits on human abilities to interact with large numbers of alters~\cite{BK1973,dunbar1998,giovanna2013}. While there are obvious constraints on the time available for interactions~\cite{giovanna2013,Milardo,Kalmijn}, additional constraints may also arise through limits on memory capacity~\cite{BK1973,Miller} or other cognitive abilities~\cite{StillerDunbar,powell2012}.

Irrespective of the specific mechanisms that act to constrain ego networks, it is reasonable to ask whether such mechanisms shape these networks in similar ways under different circumstances, giving rise to some characteristic features that persist over time despite network turnover. Here, we explore this question with a detailed analysis of the communication patterns within ego networks in an empirical setting that results in large membership turnover and changes in the closeness of relationships. In particular, we focus on the way that egos divide their communication efforts among alters, and how persistent the observed patterns are over time. We call these patterns, which may be expected to vary across individuals, social signatures. 

Over the last decade, research on human communication has been given a significant boost by the widespread adoption of new communication technologies.  The popularity of communication channels such as mobile phones and online environments has made it possible to capture micro-level quantitative data on human interactions automatically, in a way that circumvents biases inherent in retrospective self-reports~\cite{Lazer09}. However, studies using electronic communication sources typically lack information on the nature of the social relationships~\cite{onnela2007sat,palla,lambiotte,Eagle10,giovanna2013}, whereas the challenge in using survey data alone has been that these give detailed information about the nature of the social relationships, but lack quantitative information about the actual patterns of communication~\cite{oswald}. Further, in surveys the respondent burden from recording communication events with their entire ego network is very high ~\cite{fu} and people's accuracy in recalling detailed communication events is known to be limited~\cite{bernard82}.

We combine detailed, auto-recorded data from mobile phone call records with survey data. These were collected during a study~\cite{roberts2011b} which tracked changes in the ego networks of 24 students over 18 months as they made the transition from school to university or work (for details, see Materials and Methods). These changes in personal circumstances result in a period of flux for the social relationships of the participants, with many alters both leaving and entering their networks. This provides a unique setting for studying network-level structure and its response to major changes in social circumstances. This dataset combines detailed data on communication patterns from mobile phone call records with questionnaire data that explore participants' own perceptions of the quality of the relationships with all the members of their network.  More importantly, call record data contain complete time-stamped records of all calls made by an ego to alters in their network, rather than just a subset of calls an ego makes to alters who happened to be on the same mobile network as them (as has usually been the case in previous work, e.g.~\cite{onnela2007sat}). The questionnaires that augmented the call records provide information on the networks of participants that includes assessment of emotional closeness, time between face-to-face contact, and the phone numbers of alters. This allowed the call records of alters with several phone numbers (mobile phones, landlines) to be merged, giving a more accurate picture of communication between two individuals than that based on mobile phone calls alone.

These data enable us to uncover changes in the structure of the ego networks of the participants, reflected in their communication behaviour. We find a consistent pattern that is seen to be persistent over time even when there is large network turnover. This social signature  is consistent with previously observed patterns of social network site usage~\cite{marlow2009,arnaboldi2013} and text messaging~\cite{reid2006,wu2010} in that a high proportion of communication is focused on a small number of alters. A detailed analysis of the social signatures of individual participants reveals that there is individual variation in the exact way their limited communication time is allocated across their network members. Although individual signatures show some response to network turnover, they surprisingly retain much of their distinctive variation over time despite this turnover.

\section{Results}

\begin{table*}[t]
\caption{Multilevel regression models show that the number of calls made to alters significantly predicts emotional closeness to alters.}
\begin{center}
\begin{tabular*}{\hsize}{@{\extracolsep{\fill}}llll}
Parameters\tablenote{Parameter estimate standard errors listed in parentheses. PRV is the Proportional Reduction in Variance, 2LL twice the log likelihood.  $^*$ $p<0.05$, $^{**}$ $p<0.01$, $^{***}$ $p<0.001$} & Model 1 (time $t_1$) & Model 2 (time $t_2$) & Model 3 (time $t_3$) \\
\hline
\em{Regression coefficients (fixed effects)} & & &  \\
Intercept & 6.13 (0.20)$^{***}$ & 5.47 (0.42)$^{***}$ & 6.34 (0.41)$^{***}$ \\
\textbf{Number of calls (log)} & \textbf{1.63 (0.19)$^{***}$} & \textbf{1.62 (0.22)$^{***}$} & \textbf{1.35 (0.29)$^{***}$} \\
\em{Variance components (random effects)} & & & \\
Residual & 3.23 (0.26)$^{***}$ & 5.47 (0.43)$^{***}$ & 4.08 (0.42)$^{***}$ \\
Intercept & 0.42 (0.20)$^*$ & 3.21 (1.09)$^{**}$ & 2.82 (1.12)$^*$ \\
Slope & 0.08 (0.13) & & 0.30 (0.50) \\
\em{Model summary} & & & \\
Deviance statistic (-2LL) & 1436.351 & 1649.51 & 1008.18 \\
No. of estimated parameters & 5 & 4 & 5 \\
PRV & 0.27 & 0.39 & 0.38
 \\
\hline
\end{tabular*}
\end{center}
\end{table*}

\begin{figure*}[t]
\begin{center}
\includegraphics[width=0.60\linewidth]{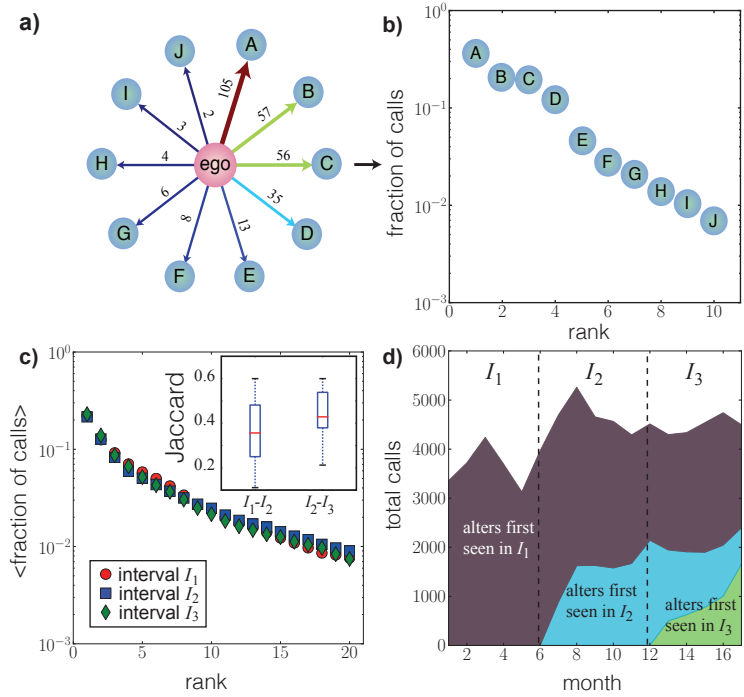}
\caption{Social signatures and network turnover. Social signatures are constructed for each ego by counting the number of calls to each of their alters, ranking the alters based on this number, and then calculating the fraction of calls to the alter of each rank (a,b). When the signatures are averaged over the set of participants for three consecutive 6-month intervals (c), it is seen that their shape is invariant although the personal networks display high turnover as indicated by the Jaccard indices between sets of 20 top ranking alters in consecutive intervals (c, inset). The network turnover is also clearly visible in (d) that shows the total numbers of calls by the participants to their alters, divided between alters that have for the first time appeared in their networks in each of the intervals.}
\label{fig:signatures}
\end{center}
\end{figure*}

\subsection{Emotional closeness of ego-identified relationships and calling behavior}
Since the main results of this study refer to the egos' social signatures which are determined on the basis of phone records with no information about the content of the conversations, we first determine whether such signatures represent an acceptable proxy for the egos' own assessments of their relationships with the alters with whom they are communicating. To this end, we employ data from questionnaires on the active personal networks of participants that were completed at three points in time:   at the beginning of the study ($t_1$), at 9 months ($t_2$) and at 18 months ($t_3$). In these surveys, egos rated their alters (kin and friends/acquaintances) for emotional closeness and number of days since last face to face encounter. We use these self-rated estimates of relationship quality to determine whether the call data reflect egos' own views of their relationships. 

To characterize egos' self-reporting of their relationships with alters we first use emotional closeness scores (scale of 1 to 10, with 1 the least and 10 the most emotionally intense relationship). Ego network data have a nested structure, where alters are clustered within participants, and thus alters cannot be treated as independent data points. We therefore used multilevel modelling, a modified form of multiple linear regression designed to deal with data with a hierarchical clustering structure~\cite{bryk} (for details see SI Appendix). In Table 1 we present the relationship between emotional closeness and number of calls (log transformed to reduce the effect of outliers). Our results show that the number of calls significantly predicts emotional closeness. The regression coefficients are positive: greater emotional closeness is associated with larger numbers of calls. 
Furthermore, there is a significant negative relationship between (log transformed) number of days between face to face contacts and number of calls, indicating that larger numbers of calls relate to smaller number of days between contacts (see SI Appendix). These results allow us to conclude that the phone call data provide a reliable estimate of relationship importance to ego.

\subsection{Social signatures and turnover in close relationships}

To study the changes in participants' communication patterns, we have divided the 18-month observation period into three consecutive intervals $I_1$, $I_2$, and $I_3$, of 6 months each. The participants took their final exams at school at month 4 of the study ($I_1$) and left the school; 18 out of the 24 participants subsequently went to university and the beginning of $I_2$ coincides with the beginning of their first university year.

In order to build the social signature of each ego for each of the intervals, we first count the number of calls to each of their alters (friends, acquaintances, kin) in the call records and subsequently rank the alters based on this number (see Methods). Then we calculate the fraction of calls as a function of alter rank to establish the social signatures for each ego in each interval, as illustrated in Fig.~\ref{fig:signatures}(a-b). 

For almost all survey participants, the signatures are characterized by a heavy tail that decreases slower than exponentially, as seen in Fig.~\ref{fig:signatures}c) which shows the social signatures averaged over the set of participants. A large fraction of communication is typically allocated to a small number of top-ranked alters: for female (male) participants, the fraction of calls to the top alter is on average $0.25\pm 0.08$ ($0.20 \pm 0.09$), and the fraction of calls to the top three alters is $0.48\pm 0.10$ ($0.40\pm 0.12$). This is in line with earlier observations in static settings, where the numbers of calls and text messages have been aggregated over some fixed time window~\cite{reid2006, wu2010, marlow2009}, as well as for the frequency of face-to-face encounters~\cite{sutcliffe2012}. It is clear that this characteristic shape of the social signatures shows considerable stability in time (see Fig.~\ref{fig:signatures}) and the signatures retain their characteristic shape despite the large turnover in network membership (Fig.~\ref{fig:signatures}).

To quantitatively measure the level of turnover in the networks of each ego, we compare the sets of alters comprising their networks in two consecutive intervals using the Jaccard index (see Methods). The high level of turnover is clearly indicated by the low average values of the Jaccard indices between successive ego networks: for the entire networks of participants, $\langle J(I_1,I_2)\rangle =0.22\pm 0.09$ and $\langle J(I_2,I_3)\rangle =0.27\pm 0.09$. The largest turnover is between intervals $I_1$ and $I_2$ ($\langle J(I_1,I_2)\rangle <\langle J(I_2,I_3)\rangle $ with $p=0.001$, paired $t$-test); this coincides with the participants finishing school in $I_1$ and the subsequent major transition in their social circumstances. Focusing on the closest relationships, the Jaccard indices for the top 20 ranking alters are  $\langle J(I_1,I_2)\rangle =0.36\pm 0.13$ and $\langle J(I_2,I_3)\rangle =0.44\pm 0.10$ (see (Fig.~\ref{fig:signatures}c, inset)), and for the top 5 ranking alters $\langle J(I_1,I_2)\rangle = 0.39 \pm 0.23$ and $\langle J(I_2,I_3)\rangle = 0.39 \pm 0.22$. Note that here, turnover includes alters dropping in rank below the top 20 or top 5. 

\begin{figure*}[t]
\begin{center}
\includegraphics[width=0.70\linewidth]{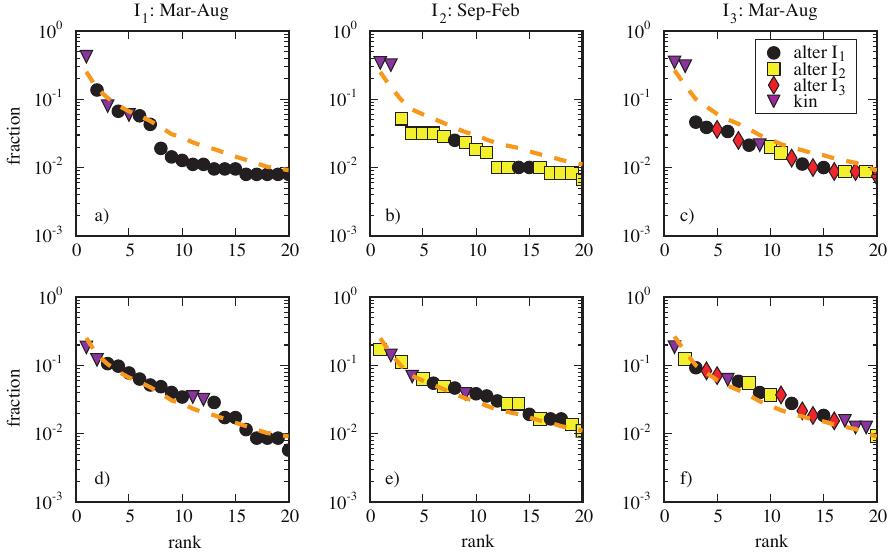}
\caption{Individual-level variation in social signatures and their evolution. The top row (panels a to c) and the bottom row (panels d to f) depict the time evolution of the social signatures of two different male participants who both went to university in another city. The symbols correspond to alters observed for the first time in intervals $I_1$ (circles), $I_2$ (squares), and $I_3$ (diamonds), or to kin (triangles) as reported by the egos. The large turnover in the networks of the participants is clearly visible. The dashed line indicates the social signature averaged over all 24 egos. In the social signatures depicted in the top row, two kin alters receive a higher-than-average fraction of outbound calls, whereas the signatures in the bottom row do not deviate much from the average. In both cases, this individual-level variation persists through all time intervals.}
\label{fig:two_signatures}
\end{center}
\end{figure*}
 
The above results indicate that whilst the turnover is the largest among lower-ranking alters, there is nevertheless significant turnover even at the level of top-ranked alters. Over the entire cohort, new alters enter the signatures at all ranks during both time points $I_2$, and $I_3$ with considerably high likelihood: the fraction of new alters entering at $I_2$ within all top 20 ranked alters is 41$\%$ (21$\%$ for $I_3$). For the top 5, the corresponding fraction is 29$\%$ for $I_2$ and 12$\%$ for $I_3$. Besides alters entering and leaving the networks, there is considerable dynamics in the ranks of alters: even for the top-ranked alters, only 42$\%$ retain their specific rank from $I_1$ to $I_2$ (54$\%$ from $I_2$ to $I_3$). For a more detailed overview of turnover and rank dynamics, see SI Appendix.

\subsection{Persistence of individual social signatures}

\begin{figure*}
\begin{center}
\includegraphics[width=0.99\linewidth]{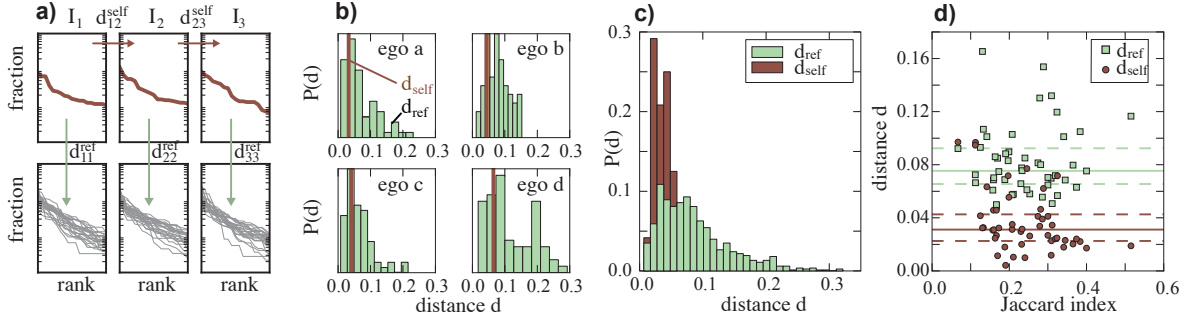}
\caption{Evidence for the persistence of social signatures at the individual level, in terms of distances between the shapes of signatures. a) A schematic of how the distances between signatures are calculated, based on Jensen-Shannon divergences. For the focal ego (top row), self-distances  $d_\mathrm{self}$ are calculated for signatures in consecutive intervals and averaged. Reference distances $d_\mathrm{ref}$  are calculated for each interval between the signatures of the focal ego and all other egos (bottom row). These are averaged over the three intervals for each pair of egos (focal, other). Panel b) shows the values of the average self-distances  $d_\mathrm{self}$ and histograms for reference distances  $d_\mathrm{ref}$ for four sample egos, indicating that the shapes of each ego's signatures in consecutive intervals are typically more similar than they are to those of other egos. Panel c) displays the distributions of all self-distances $d_\mathrm{self}$ and reference distances $d_\mathrm{ref}$, for all egos, verifying the larger similarities between egos' signatures in consecutive intervals. d) The self-distances $d_\mathrm{self}$ show a moderate level of correlation ($r$=-0.41, $p$=0.0034) with turnover as measured with the Jaccard index $J$, but nevertheless mostly remain below reference distances. The scatter plot shows the coordinate pairs 
$\left\{ J_{12}(i),d_{12}^\mathrm{self}(i)\right \}$ and $\left\{J_{23}(i),d_{23}^\mathrm{self}(i)\right\}$ 
for each ego $i$ (circles). For comparison, reference distances are also displayed as the coordinate pairs  
$\left\{J_{12}(i),d_{22}^\mathrm{ref}(i)\right\}$ 
and 
$\left\{J_{23}(i),d_{33}^\mathrm{ref}(i)\right\}$. Solid and dashed lines denote the median and the quartiles.
}
\label{fig:distances}
\end{center}
\end{figure*}

The above analysis indicates that the pattern in which top-ranked alters receive a substantial fraction of communication remains persistent over time, even as those alters change. We next turn to characterizing the stability of individual social signatures of egos in more detail. Fig.~\ref{fig:two_signatures} illustrates the evolution of two of these signatures for the three intervals, with colors indicating alters that enter their networks at different times. Both signatures retain much of their shape despite network turnover. 

For quantifying the similarity of signatures, we introduce the Jensen-Shannon divergence (JSD, see Methods) as a measure of distance (shape difference) between of two signatures. The aim is to investigate the similarity of the shape of an ego's social signature in different time intervals, using the distances to the signatures of all other egos as a reference. To this end, for each ego $i$, we first calculate JSD values $d$ for the two pairs of social signatures in consecutive time intervals $(I_1,I_2)$ and $(I_2,I_3)$, and average over these two values in order to obtain the ego's self-distance $d_\mathrm{self}(i)$. Denoting the JSD distances between the social signatures of egos $i$ and $j$ in intervals $a$ and $b$ by $d_{ab}^{ij}$, we first calculate ego $i$'s self-distances  between intervals $(I_1,I_2)$ and $(I_2,I_3)$ as $d_{12}^\mathrm{self}(i)=d_{12}^{ii}$ and $d_{23}^\mathrm{self}(i)=d_{23}^{ii}$. Then we average over these to get the self-distance $d_{\mathrm{self}}(i)=\frac{1}{2}\left[d_{12}^\mathrm{self}(i)+d_{23}^\mathrm{self}(i)\right]$ (see Fig.~\ref{fig:distances}a). Low values of $d_\mathrm{self}$ then denote a high similarity between the ego's signatures in consecutive intervals.
 For reference, we calculate the distances between the signatures of the focal ego $i$ and those of all other egos $j$  for each of the three intervals, e.g. for focal ego $i$, interval $I_1$ and reference ego $j$, the reference distance $d_\mathrm{ref}$ is $d_{11}^{ij}$. 
 


The results in Fig.~\ref{fig:distances} clearly indicate that on average, the shapes of the social signatures of participants show a tendency to persist in time, as the distances between one participant's consecutive signatures $d_\mathrm{self}$ are on average much lower than the reference distances to the signatures of other participants $d_\mathrm{ref}$. Fig.~\ref{fig:distances}b displays the self-distances and distributions of reference distances to the signatures of all other participants for four example egos, and Fig.~\ref{fig:distances}c shows the distributions of all self-distances $P(d_\mathrm{self})$ and reference distances $P(d_\mathrm{ref})$.
On average, for each ego, $82\%\pm 12\%$ of the distances to others $d_\mathrm{ref}$ were greater than $d_\mathrm{self}$ . 
 Averaged over all egos, the average self-distance is  $\langle d_\mathrm{self}\rangle =0.036\pm 0.014$ while the average distance to other egos is  
 $\langle d_\mathrm{ref}\rangle =0.086\pm 0.055$  ($\langle d_\mathrm{self}\rangle <\langle d_\mathrm{ref}\rangle $  with $p<10^{-4}$ , Welch's $t$-test). For verification, we used the  $\ell^2$ -norm as an alternative distance measure (see Methods); this yields a qualitatively similar outcome $\langle d_\mathrm{self}\rangle =0.096\pm 0.039$,  $\langle d_\mathrm{ref}\rangle =0.154\pm 0.084$, $\langle d_\mathrm{self}\rangle <\langle d_\mathrm{ref}\rangle $  with $p<10^{-4}$.

Although the individual signatures retain much of their shape, there is nevertheless some variation in their persistence, as seen in the distribution of self-distances in Fig.~\ref{fig:distances} c). One plausible candidate explanation for this is the level of turnover in the egos' networks. To determine if turnover in relationships has an effect on the persistence of signatures, we examined the relationship between the self-distances and the Jaccard indices. For each ego $i$, we determined the Jaccard indices for the two pairs of intervals $J_{12}(i)=J(I_1,I_2;i)$ and $J_{23}(i)=J(I_2,I_3;i)$ and the corresponding self-distances $d_{12}^\mathrm{self}(i)$ and $d_{23}^\mathrm{self}(i)$. Fig.~\ref{fig:distances}d shows a scatter plot of these $N=48$ pairs of values (circles), indicating a moderate level of correlation (Pearson correlation coefficient $r$=-0.41, $p$=0.0034). However, even with this variation, most signatures retain their distinctive characteristics, as the scatter of the self-distance values is systematically below the scatter of the reference distances 
$d_{22}^\mathrm{ref}(i)=\frac{1}{N_\mathrm{egos}-1}\sum_{j \neq i} d_{22}^{ij}$ and $d_{33}^\mathrm{ref}(i)=\frac{1}{N_\mathrm{egos}-1}\sum_{j \neq i} d_{33}^{ij}$ 
displayed in the same plot (squares). For Jaccard indices calculated for the top 5 and top 10 alters only, the correlation is no longer significant ($r$=-0.029 with $p$=0.84 and $r$=-0.19 with $p$=0.20, respectively), indicating that turnover among the closest relationships alone does not explain the variation in self-distances.

\section{Discussion}

Our results establish three novel findings: (1) there is a consistent, broad and robust pattern in the way people allocate their communication across the members of their social network, with a small number of top-ranked, emotionally close alters receiving a disproportionately large fraction of calls; (2) within this general pattern, there is clear individual-level variation so that each individual has a characteristic social signature depicting their particular way of communication allocation; and (3) this individual social signature remains stable and retains its characteristic shape over time and is only weakly affected by network turnover. Thus individuals appear to differ in how they allocate their available time to their alters, irrespective of who these alters are. Further, our subsidiary analyses (see SI Appendix) suggest that this finding applies not just to call frequencies, because the frequency of calls to an alter correlates with emotional closeness and frequency of face-to-face interactions.

The patterns displayed in the social signatures reflect the fact that ego networks are typically layered into a series of hierarchically inclusive subsets of relationships of different quality. This pattern has been noted, for example, in shipboard networks~\cite{BK1973} as well as friendship social networks and the structure of natural human communities~\cite{Zhou2005}. Bernard and Killworth~\cite{BK1973} were perhaps the first to suggest that this structuring was likely due to a psychological (i.e.~cognitive) constraint. An alternative (but not necessarily mutually exclusive) possibility is that they arise from the fact that the time available for interaction is limited, and individuals partition their available social time to reflect their social or emotional preferences~\cite{roberts2011}. One problem is that it may be difficult to separate out the cognitive and time aspects of relationships, since it seems that the time invested in a relationship may determine its emotional quality~\cite{sutcliffe2012,roberts2011}. A rather different kind of cognitive mechanism derives from the social brain hypothesis~\cite{dunbar1998}:  recent neuroimaging studies of humans demonstrate a correlation  at the individual level between core brain regions (notably those in the prefrontal cortex) and size of the innermost layers of ego networks~\cite{powell2012,kanai}, with similar findings reported from monkeys~\cite{sallet}. More importantly, Powell et al.~\cite{powell2012} were able to show that this relationship between brain region volume and network size is mediated by forms of social cognition known as Ômentalising' (or mind-reading). Individuals' mentalising competences may limit the numbers of individuals they can maintain at any given level of emotional closeness~\cite{sutcliffe2012}. An alternative (yet not necessarily mutually exclusive) possibility is that the constraint arises from differences in personality, with some individuals preferring to have a few, intense relationships and others preferring more, less intense ones~\cite{swickert_ind_diff}. Determining the extent to which the unique pattern represented by an individual's social signature is a consequence of time, cognitive or other constraints will require more detailed study of rather different kinds of data. 

Whatever the mechanism involves, it seems that, because of these constraints, individuals cannot increase the number of alters they communicate with at maximum rate, but must downgrade (or drop) some individuals if they wish to add new ones to their preferred network at a high level of emotional intensity. As a result, the overall shape of their social signature  is preserved. Indeed, our findings can be taken to suggest that these patterns are so prescribed that even the efficiencies provided by some forms of digital communication (in this case, cellphones) are insufficient to alter them. This might explain the observation that most of the traffic on social networking sites (SNSs) is directed at a very small number of individuals~\cite{marlow2009} and that this appears to be resistant to change~\cite{pollet2011} no matter how SNSs attempt to engineer a wider circle of social contacts.   It would be particularly instructive to explore the implications of this study on large scale datasets derived from electronic media where a wider variety of personal circumstances can be explored. Even though these offer only limited kinds of data, our findings at least suggest that these would provide reasonable proxies for relationship quality.  Doing so would enable us to explore differences due to age, gender and personality, as well as the consequences of the slower pattern of network turnover that occurs naturally over time as individuals switch the focus of their social investment.

\section{Materials and methods}
\subsection{Personal network survey and call records} 
We used longitudinal data on the social networks of thirty participants (15 males and 15 females, aged 17 to 19 years old) in their last year of secondary school, collected over an 18-month period during the transition from school to university or work (for full details, see 16). Participants completed a questionnaire on their active personal network at three points in time: at the beginning of the study ($t_1$), at 9 months ($t_2$) and at 18 months ($t_3$). The analysis in this study is based on the 24 participants (12 males, 12 females) who completed all three questionnaires and used their mobile phones throughout the study. To elicit their ego network, participants were asked to list all unrelated individuals "for whom you have contact details and with whom you consider that you have some kind of personal relationship (friend, acquaintance, someone you might interact with on a regular basis at school, work or university)". The participants were also asked to list all their known relatives. For all individuals listed, participants were asked to provide both landline and mobile phone numbers. In each survey, for both kin and friends/acquaintances, the participants were asked to indicate the emotional intensity of the relationship by providing an emotional closeness score, measured on a 1-10 scale, where 10 is someone "with whom you have a deeply personal relationship".

At $t_1$, all participants lived in the same large UK city. At month 4 of the study, the participants took their final exams at school ("A-levels") and left the school. Of the 24 participants who completed all three questionnaires, six participants stayed in the home city and worked, not going to University; eight went to university in the city (which has two large universities) and the remaining 10 went to university elsewhere in England.

As compensation for participating in the study, participants were given a mobile phone, with an 18-month contract from a major UK mobile telephone operator. The line rental for the mobile phone was paid for, and included 500 free monthly voice minutes (to landlines or mobiles) and unlimited free text messages. For each participant, we obtained itemized, electronic monthly phone invoices that listed all outgoing calls (recipient phone number, time and duration of calls). The electronic PDF invoices were parsed into machine-readable form. The questionnaire data and the call dataset form the main basis for our analysis.

\subsection{Constructing ego-centric call networks}

For each participant in the study (ego), we used the list of kin and friends/acquaintances (alters) generated in response to the three social network questionnaires and combined it with the electronic phone invoices to construct a set of ego-centric call networks. If an alter was listed as having multiple phone numbers, a mobile and a fixed line number, a call by the ego to either number was recorded as a call between ego and the alter. Phone numbers appearing on the invoices but not listed in the questionnaire responses were treated as unique alters; however, service numbers (such as those with 0800 prefixes) were filtered out. All alters appearing in the phone records that were believed to be non-service numbers were used in the calculation of the signatures, independently of being recalled or not by the egos at the time the questionnaires where completed (percentages for the number of alters not recalled are directly related to call frequency; for details see SI Appendix). The 18-month observation period of electronic phone invoices was divided into three consecutive intervals of 6 months each ($I_1$: March-August, $I_2$: September-February, $I_3$: March-August). For each ego in each of the three intervals, we counted the total number of calls made to each alter. Using the alter-call-counts per interval, we ranked the egos from most called to least called, and calculated their social signatures depicting the total fraction of calls to an alter as a function of the alter's rank. 

\subsection{Analyzing social signatures}
We quantify the variation between the sets of alters an ego calls in two time intervals with the Jaccard coefficient,
\begin{equation}
J(I_i, I_j) = \frac{|A(I_i) \cap A(I_j)|}{|A(I_i) \cup A(I_j)|}
\end{equation}
where $A(I_i)$ and $A(I_j)$ are the sets of alters called by the ego in two time intervals $I_i$ and $I_j$, respectively. Then $J=1$ if the sets are the same, and $J=0$ if the sets have no common alters. For a pairwise comparison of the social signatures between two different egos or two different time intervals for a single ego we measure the Jensen-Shannon divergence (JSD)~\cite{lin1991} defined as
\begin{equation}
JSD\left(P_1,P_2\right)=H\left(\frac{1}{2}P_1+\frac{1}{2}P_2\right)-\frac{1}{2}\left[H(P_1)+H(P_2)\right],
\end{equation}
where $P_1$ and $P_2$ are the two signatures where $P_i=\left\{p_i(r)\right\}$ such that $p_i(r)$ is the fraction of calls to the alter of rank $r$ in pattern $i$. Additionally, $H(P)$ is the Shannon entropy,
\begin{equation}
H(P)=-\sum_{r=1}^k p(r)\log p(r),
\end{equation}
where $p(r)$ is as above and $k$ is the maximum rank, \emph{i.e.}~the total number of alters called. The Jensen-Shannon divergence is a generalized form of the Kullback-Leibler divergence (KLD) such that $JSD(P_1,P_2)\in[0,\infty)$, and $JSD(P_1,P_2)=0$ iff the distributions are identical. We chose JSD over KLD due
to its capacity to deal with zero probabilities $p(r)=0$. The maximum number of alters called by an ego in a given time interval, $k$, varies depending on the ego and the interval; therefore, if $k_2>k_1$ is the larger number, we assign $p_1(r_1)=0$ for $k_1>r_1\geq k_2$, \emph{i.e.}~zero-pad the series of fractions of 
calls such that they are of the same length. Additionally, for validating the pairwise comparison results, we also calculated the $\ell^2$-norm for pairs of social signatures, defined as $\ell^2=\sqrt{\sum_{r=1}^k\left|p_1(r)-p_2(r)\right|^2}$.

\begin{acknowledgments}
We acknowledge support by EU's 7th Framework Program's FET-Open to ICTeCollective project no. 238597. J.S. was supported by the Academy of Finland (grant n:o 260427). E.L. and F.R.-T. were supported by the Oxford Martin School. E.A.L. was supported by the Rockefeller Foundation. S.G.B.R. was supported by the Lucy to Language British Academy Centenary Research Project, the EPSRC/ESRC Developing Theory for Evolving Socio-Cognitive Systems (TESS) project and an EPSRC Knowledge Secondment Award. R.I.M.D.'s research is funded by an ERC Advanced grant. We thank Renaud Lambiotte for useful discussions.
\end{acknowledgments}

\appendix*

\setcounter{figure}{0} \renewcommand{\thefigure}{S\arabic{figure}}
\setcounter{table}{0} \renewcommand{\thetable}{S\arabic{table}}

\section{Supplementary Information}

\subsection{The relationship between mobile phone calls and emotional closeness}

\begin{table*} 
\caption{Summary of nine separate multilevel models predicting emotional closeness from mobile calls variables. Only the regression coefficients (fixed effects) and proportional reduction in variance (PRV) are shown for each model.}
\label{tableSI:models}
\centering
\begin{tabular*}{\hsize}{@{\extracolsep{\fill}}lll}
Parameters & Regression coefficient\tablenote{ $^{***}$ $p<0.001$} & PRV \\
\hline
\em{Alter rank based on number of calls (log)} & & \\
 \hspace{0.25cm}Alter rank ($t_1$) & -2.08 (0.27)$^{***}$ & 0.29 \\
 \hspace{0.25cm}Alter rank ($t_2$) & -2.19 (0.33)$^{***}$ & 0.39 \\
 \hspace{0.25cm}Alter rank ($t_3$) & -1.95 (0.44)$^{***}$ & 0.37 \\
\em{Durations of calls (log} & & \\
 \hspace{0.25cm}Duration ($t_1$) & 0.84 (0.11)$^{***}$ & 0.19 \\
 \hspace{0.25cm}Duration ($t_2$) & 0.87 (0.17)$^{***}$ &  0.39 \\
 \hspace{0.25cm}Duration ($t_3$) & 0.88 (0.15)$^{***}$ & 0.39 \\
\em{Alter rank based on duration of calls (log)} & & \\
 \hspace{0.25cm}Alter rank ($t_1$) & -1.54 (0.23)$^{***}$ & 0.19 \\
 \hspace{0.25cm}Alter rank ($t_2$) & -2.02 (0.34)$^{***}$ & 0.38 \\
 \hspace{0.25cm}Alter rank ($t_3$) & -1.97 (0.34)$^{***}$ & 0.39 \\
\hline
\end{tabular*}
\end{table*}

We examined the relationship between mobile phone calls and the emotional closeness ratings given by egos to each alter. Personal network data have a nested structure, where alters are clustered within participants, and thus alters cannot be treated as independent data points. In this analysis, we therefore used multilevel modelling, a modified form of multiple linear regression designed to deal with data with a hierarchical clustering structure~\cite{Bryk}. We used a two-level model, in which alters (Level 1) were clustered within the 24 egos (Level 2). In all the models, the dependent variable was emotional closeness, measured on a 1-10 scale (with 10 being very close). For each alter, emotional closeness was reported at three time points -- at the beginning of the study ($t_1$), after 9 months ($t_2$) and at 18 months ($t_3$). Thus for each set of call data, separate models were constructed for $t_1$, $t_2$ and $t_3$. For the call data, we used the two months of calls around the date of the questionnaire, in order to examine if the emotional closeness rating given to each alter was related to their calling pattern around the time questionnaire was being completed. 

To allow for comparison across the different models, we attempted to fit all models with the same structure, with a fixed effect for the call data, random intercepts and random slopes. However, in some models, fitting the random slopes caused the models not to converge, and thus these models were calculated without including random slopes. The proportional reduction in variance (PRV) between an empty model and the final model was used to estimate the local effect size~\cite{Peugh,Carson}. All analysis for the multilevel models was carried out using IBM SPSS Statistics 20.

In the models, we examined the relationship between the emotional closeness ratings and: i) Number of calls; ii) Rank of alters based on number of calls; iii) Duration of calls; iv) Rank of alters based on duration of calls. To reduce the effects of outliers on the model, all the call data (number of calls, duration of calls and rank) were log transformed.

The model for emotional closeness and number of calls is given in full in the main text. The results of the other models are summarised in Table S1. The key result is that in all the models the level of emotional closeness to alters was significantly predicted by mobile phone calls, or alter rank based on calls. For calls, the regression coefficients were positive, indicating that greater numbers/longer durations of calls were associated with higher levels of emotional closeness. For ranks, the regression coefficients were negative, indicating that the top ranked alters (i.e. those with low rank numbers) had higher levels of emotional closeness. Thus egos were emotionally close to the top ranked alters that received a large number or long duration of phone calls. The PRV indicates that the models explained between 19 and 39\% of the variance in emotional closeness ratings.

\subsection{The relationship between mobile phone calls and face-to-face contact}

\begin{table*} 
\caption{Summary of twelve separate multilevel models predicting days to last face-to-face contact (log) from mobile calls variables. Only the regression coefficients (fixed effects) and proportional reduction in variance (PRV) are shown for each model.}
\label{tableSI:models2}
\centering
\begin{tabular*}{\hsize}{@{\extracolsep{\fill}}lll}
Parameters & Regression coefficient\tablenote{$^*$ $p<0.05$ $^{**}$ $p<0.01$ $^{***}$ $p<0.001$} & PRV \\
\hline
\em{Number of calls (log)} & & \\		
 \hspace{0.25cm}Number of calls ($t_1$) &	-0.75 (0.14)$^{***}$	& 0.18 \\
  \hspace{0.25cm}Number of calls ($t_2$) &	-0.70 (0.16)$^{***}$	& 0.13 \\
  \hspace{0.25cm}Number of calls ($t_3$) &	-1.10 (0.20)$^{***}$	& 0.14 \\
\em{Alter rank based on number of calls (log)} & & \\		
    \hspace{0.25cm}Alter rank ($t_1$) &	0.94 (0.17)$^{***}$	& 0.18 \\
    \hspace{0.25cm}Alter rank ($t_2$) &	0.82 (0.22)$^{***}$	& 0.12 \\
    \hspace{0.25cm}Alter rank ($t_3$) &	1.18 (0.26)$^{***}$	& 0.12 \\
\em{Durations of calls (log)} & & \\		
    \hspace{0.25cm}Duration ($t_1$) &	-0.12 (0.09)	& 0.10 \\
    \hspace{0.25cm}Duration ($t_2$) &	-0.29 (0.10)$^{**}$	& 0.10 \\
    \hspace{0.25cm}Duration ($t_3$) &	-0.53 (0.12)$^{***}$	& 0.09 \\
\em{Alter rank based on duration of calls (log)} & & \\		
    \hspace{0.25cm}Alter rank ($t_1$) &	0.18 (0.18)	& 0.10 \\
    \hspace{0.25cm}Alter rank ($t_2$) &	0.45 (0.22)$^{*}$	& 0.09 \\
    \hspace{0.25cm}Alter rank ($t_3$) &	0.86 (0.25)$^{**}$	& 0.08 \\
\hline
\end{tabular*}
\end{table*}

The relationship between mobile phone calls and the frequency of face-to-face contact was investigated using exactly the same procedure as described above. In the questionnaires, participants were asked how many days ago they last made face-to-face contact with each alter~\cite{roberts2011b}. This variable was log transformed to reduce the effect of outliers. The relationship between face-to-face contact and the different mobile call variables (call number, alter rank based on call number, call duration, alter rank based on call duration) was investigated for each of the three time periods, using separate multilevel models for each variable and time period, resulting in 12 separate models. The model fitting procedure was the same as described above, but for all models fitting random slopes caused the models not to converge, so all models were fitted with random intercepts only.

The results are summarized in Table S2. In all models (except two) face-to-face contact was significantly predicted by mobile phone calls, or alter rank based on calls. For the number and duration of calls, the regression coefficients were negative, indicating that a smaller number of days to last face-to-face contact was associated with a greater number or longer duration of calls. For alter ranks, the coefficients were positive, indicating that the top ranked alters (i.e. those with low rank numbers) had a smaller number of days to last face-to-face contact. Thus egos were in frequent face-to-face contact with the top ranked alters that received a large number or long duration of phone calls. The PRV indicates that the models explained between 8 and 18\% of the variance in days to last face-to-face contact, which was lower than for the models predicting emotional closeness.

\subsection{Using call durations instead of call numbers when ranking alters}

There are two possible ways of defining the weights of relationships between egos and alters on the basis of call records. One option is to use the total number of calls to each alter as weights for ranking the alters. The results in the main text are based on this approach. Alternatively, one can use the total call duration, that is, the sum of durations of all calls between an ego and an alter in a given time interval. Both types of weights can be argued to reflect the intensity of the relationship -- often, they are highly correlated (see, e.g.,~\cite{Onnela2}). This is the case for the present data, too, and as shown below, the results of this article are not qualitatively changed if weights based on call durations are used.

Figure S1 displays a scatter plot of both types of weights for all ego-alter pairs, for each of the three 6-month intervals $I_1$, $I_2$, and $I_3$. Call durations and numbers are clearly seen to correlate. This is verified by calculating the Pearson correlation coefficients for the weights. When the Pearson correlation coefficients $r$ between the two types of weights are calculated separately for each ego's set of alters and each interval, and then averaged over all egos, one obtains for the three intervals $\langle r(I_1) \rangle=0.80 \pm 0.15$ (std), $\langle r(I_2) \rangle=0.78 \pm 0.13$ (std), and $\langle r(I_3) \rangle=0.80 \pm  0.17$ (std).

To further verify that the results are not affected by the choice of weights, we have repeated all analysis using call durations as weights. The results are summarized in Figure S2. With call duration weights, for female (male) participants, the fraction of time spent talking on the phone with the top alter is on average $0.36 \pm 0.16$ $(0.28 \pm 0.13)$, and the fraction of time spent on the phone with the the top three alters is $0.61 \pm 0.16$ $(0.48 \pm 0.15)$. Measuring network turnover with the average values of the Jaccard indices between the personal networks of egos, one obtains for the top 20 
 ranking alters of each ego's network $\langle J(I_1, I_2)\rangle=0.33 \pm 0.14$ and $\langle J(I_2, I_3\rangle=0.39 \pm 0.12$ (see Fig S2, panel a) inset).
 Measuring the differences between signature patterns using the Jensen-Shannon   divergence, on average, for each ego, $81\% \pm 14\%$ of the distances to others $d_\mathrm{ref}$ were greater than $d_\mathrm{self}$.  Averaged over all egos, the average self-distance is  $\langle d_\mathrm{self}\rangle =0.045\pm 0.023$ while the average distance to other egos is  $\langle d_\mathrm{ref}\rangle =0.122 \pm 0.094$ ($\langle d_\mathrm{self}\rangle$ is smaller than $\langle 
 d_\mathrm{ref}\rangle$ with $p<10^{-4}$, Welch's $t$-test). Again, for verification, we used the $\ell^2$-norm as an alternative distance measure; this yields a qualitatively similar outcome $\langle d_\mathrm{self}\rangle=0.15 \pm 0.06$,  $\langle d_\mathrm{ref}\rangle=0.25 \pm 0.15$, ($\langle d_\mathrm{self}\rangle$ is smaller than $\langle d_\mathrm{ref}\rangle$ with $p<10^{-4})$.
  
\subsection{Alters not recalled in surveys}

All participants completed a questionnaire on their active personal network at three points in time: at the beginning of the study ($t_1$), at 9 months ($t_2$) and at 18 months ($t_3$). These questionnaires provided us with information on mobile and landline telephone numbers associated with alters and the self-assessed emotional closeness with alters. We have used this information in establishing the relationship between call frequency and emotional closeness, as well as in mapping phone numbers to alters for alters with multiple listed phone numbers. However, there is a large fraction of phone numbers in the electronic invoices that do not appear in survey results. Note that for establishing the social signatures and network turnover figures, we have used data on all calls from the electronic invoices, whether the numbers are listed in surveys or not.

We call the alters whose phone numbers are listed in invoices but who do not appear in survey responses non-recalled alters and denote their fraction of all alters by $f$. For each of the three intervals, the fractions of non-recalled alters averaged over all egos are $f(I_1)=0.63 \pm 0.21$ (std), $f(I_2)=0.68 \pm 0.18$ (std), and $f(I_3)=0.70 \pm 0.17$ (std). 

Non-recalled alters are not evenly distributed among all alters, but there is a bias towards low-frequency/low-duration links. This is seen in the lower fractions $f_C$ of calls to non-recalled alters. The values for each of the three intervals are $f_C(I_1)=0.33 \pm 0.21$, $f_C(I_2)=0.39 \pm 0.19$ and $f_C(I_3)=0.45 \pm 0.20$. Likewise, when calculating the probability density distributions (PDFs) separately for the weights of recalled/non-recalled ego-alter pairs, it is seen that the averages are much lower for the non-recalled alters (Figure S3). Note that in Figure S3 the weights have been separately normalized to unity for each ego before calculating the PDFs for the link weights, as the overall call activity levels of egos are very different. The distributions for recalled alters are clearly shifted towards larger values. For numbers of calls as weights, the average ego-normalized weights for recalled alters are $w(I_1)=0.032$, $w(I_2)=0.025$, and $w(I_3)=0.028$ while for non-recalled alters $w(I_1)=0.005$, $w(I_2)=0.005$ and $w(I_3)=0.006$.
 
\subsection{Ranks and network turnover}

The level of network turnover shows some dependence on ranks -- see Figures S4 and S5 that show the rank-specific fractions of alters leaving the egos' networks and entering the ego's networks for the first time. There are more low-ranked alters than high-ranked alters leaving the egos' networks, and new alters are likely to make their entry at low ranks. This is as expected; it would be rather surprising if high-ranked, emotionally close alters were replaced with newcomers as easily and as frequently as low-ranked alters.   

Nevertheless, there is significant turnover even among the highest ranks, and a substantial fraction of alters who were not present in the first interval enter the highest ranks in $I_2$ (Figure S5). Moreover, a large fraction of high-rank alters do not retain their ranks, as seen in Figure S6 that depicts the fraction of alters of each rank that retain their rank across intervals: even of the highest-ranked alters, only 42\% retain their top rank from interval $I_1$ to $I_2$, and 54\% from $I_2$ to $I_3$. Finally, to illustrate the simultaneous effects of network turnover and rank dynamics, Figure S7 shows the changes of the states of alters of given ranks for each of the 24 egos, across the two pairs of intervals  $I_1$ to $I_2$ and $I_2$ to $I_3$.

\begin{figure}
\begin{center}
\includegraphics[width=0.6\linewidth]{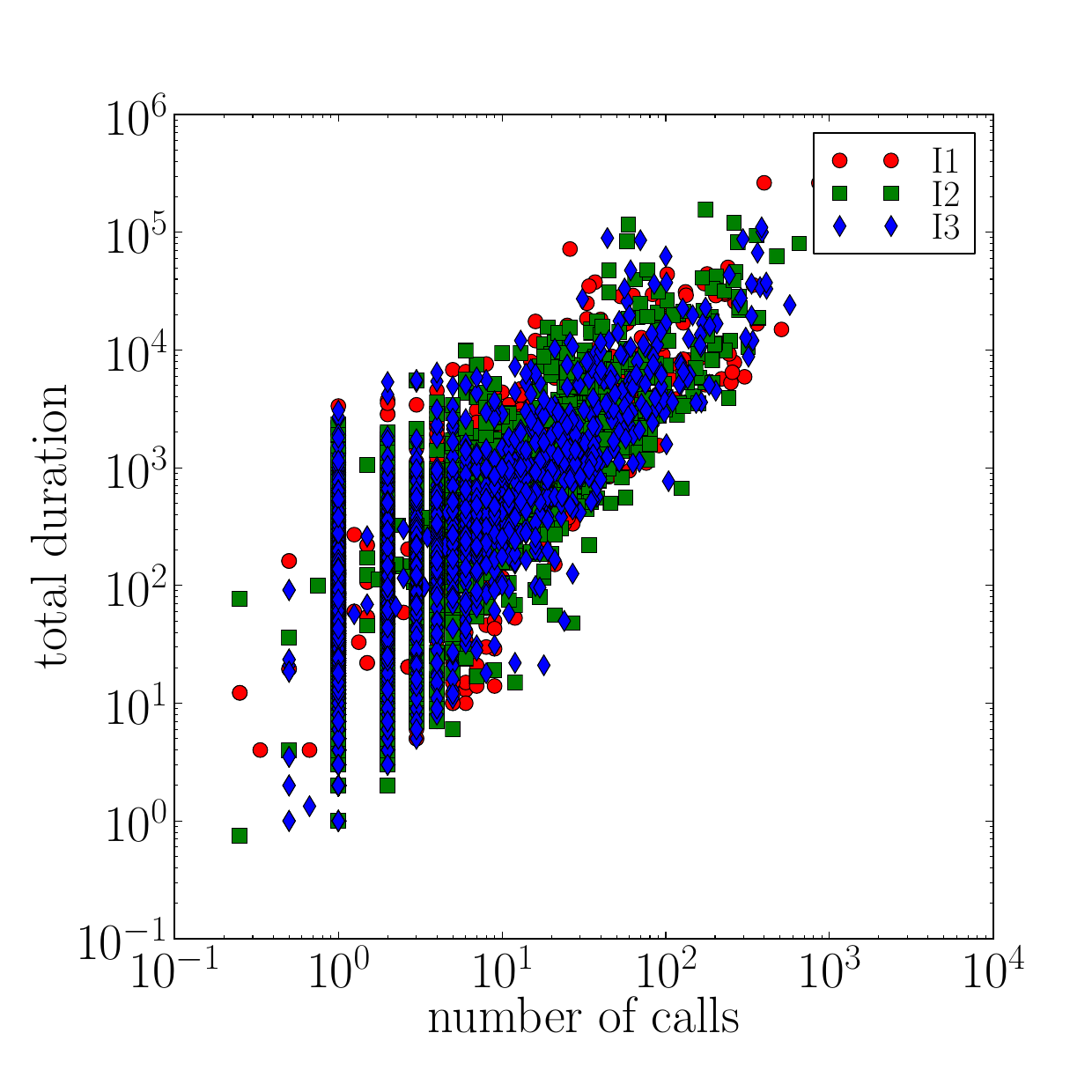}
\caption{The relationship between weights representing numbers of calls and total call durations (in seconds) for each ego-alter pair, separately for the three 6-month intervals.}
\label{Fig_S1}
\end{center}
\end{figure}

\begin{figure}
\begin{center}
\includegraphics[width=0.9\linewidth]{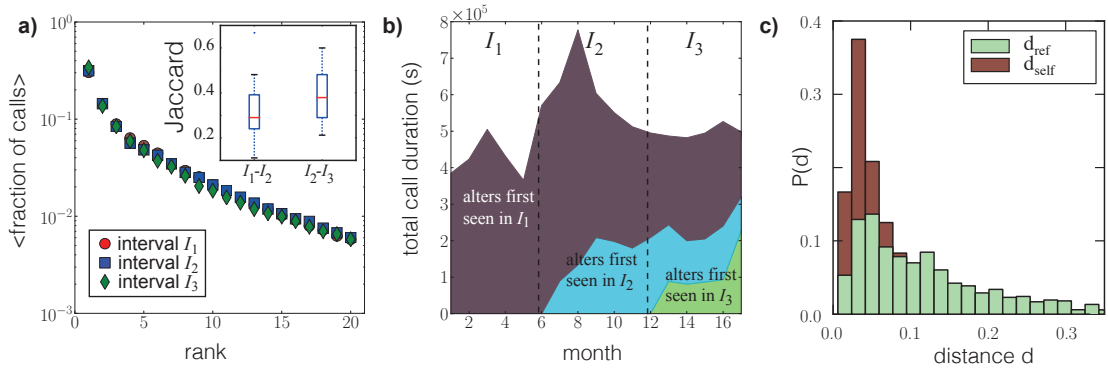}
\caption{Main results with total call durations as weights instead of call numbers. a) The averaged signature patterns for each of the three intervals and the Jaccard indices between sets of top 20 alters. b) Total call durations as a function of time, divided among alters based on their first time of appearance in the networks. c) Distribution of self and reference distances between signature patterns.}
\label{Fig_S2}
\end{center}
\end{figure}

\begin{figure}
\begin{center}
\includegraphics[width=0.9\linewidth]{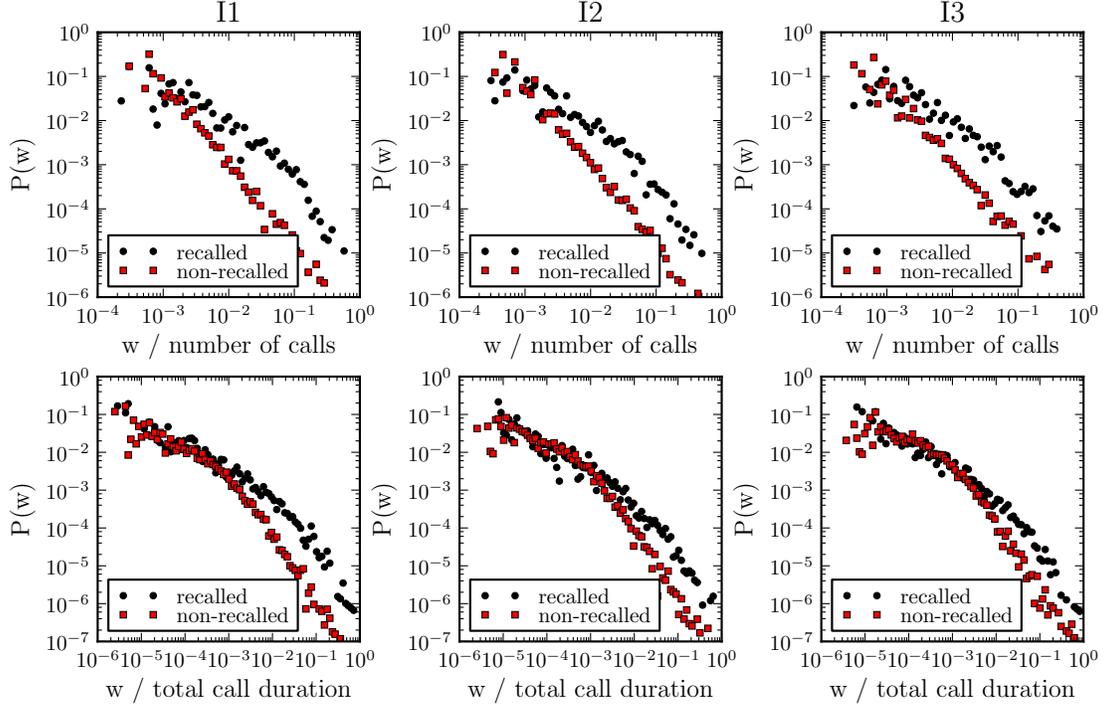}
\caption{Probability density distributions for the ego-alter pair weights (top row: number of calls defines weight, bottom row: total call duration defines weight), separately for recalled alters and non-recalled alters and for each 6-month interval (columns). Weights have been normalized separately for each ego. Recalled alters are on average associated with higher weights; please note that a logarithmic scale has been used.
}
\label{Fig_S3}
\end{center}
\end{figure}

\begin{figure}
\begin{center}
\includegraphics[width=0.8\linewidth]{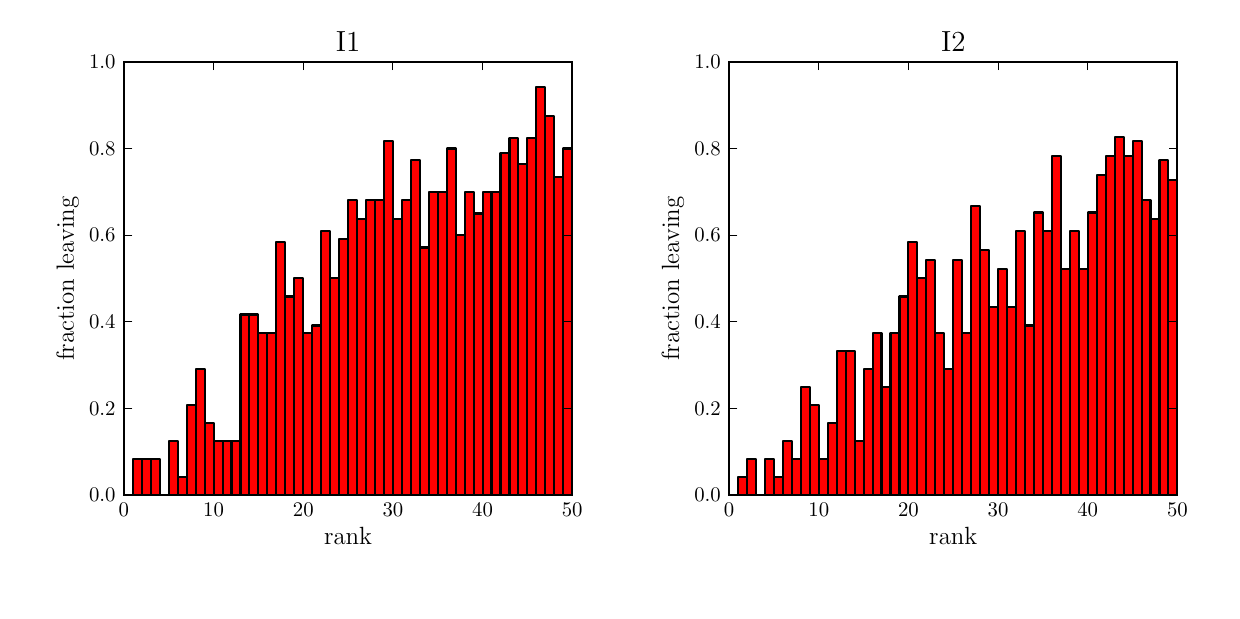}
\caption{Fraction of alters who subsequently left the egos' personal networks, as a function of rank. Left panel displays alters active in $I_1$ but no longer in the personal networks in $I_2$, and the right panel alters who were present in $I_2$ but left before $I_3$. The lower the rank of an alter, the more likely that alter is to leave the network.
}
\label{Fig_S4}
\end{center}
\end{figure}

\begin{figure}
\begin{center}
\includegraphics[width=0.8\linewidth]{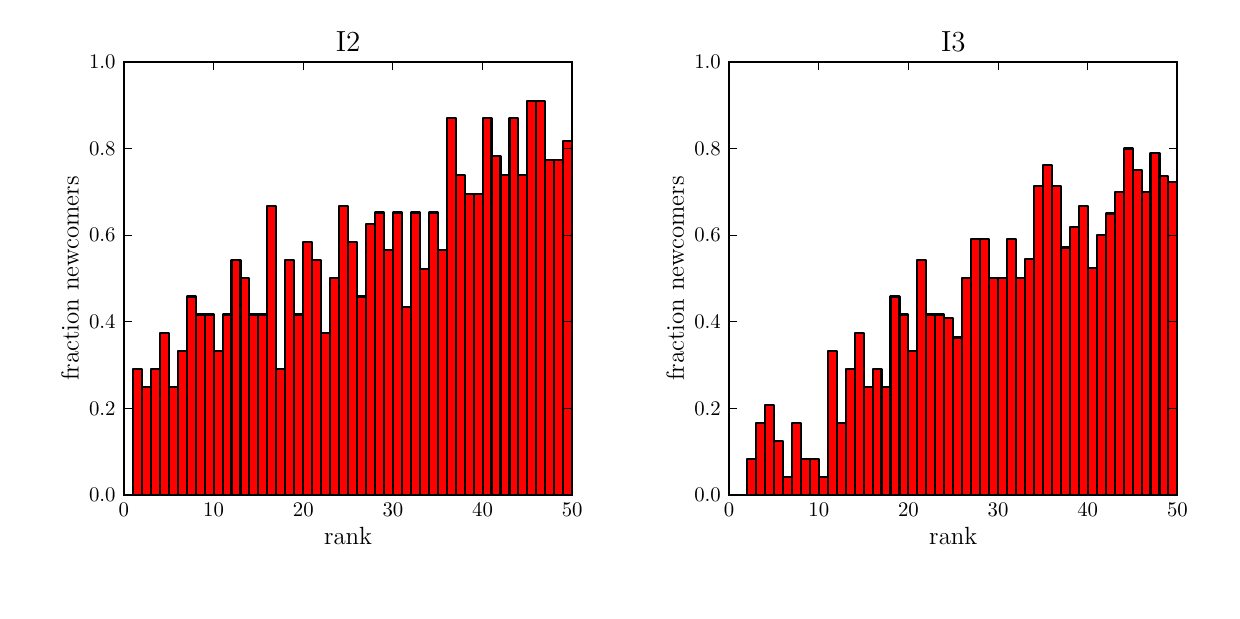}
\caption{Fraction of alters who are new to the egos' personal networks, as a function of rank. Left panel displays alters who entered the networks in $I_2$, and the right panel alters entered in $I_3$. Although the fraction of newcomers is highest for low ranks, it is not negligible even for the highest ranks, especially for $I_2$.
}
\label{Fig_S5}
\end{center}
\end{figure}

\begin{figure}
\begin{center}
\includegraphics[width=0.8\linewidth]{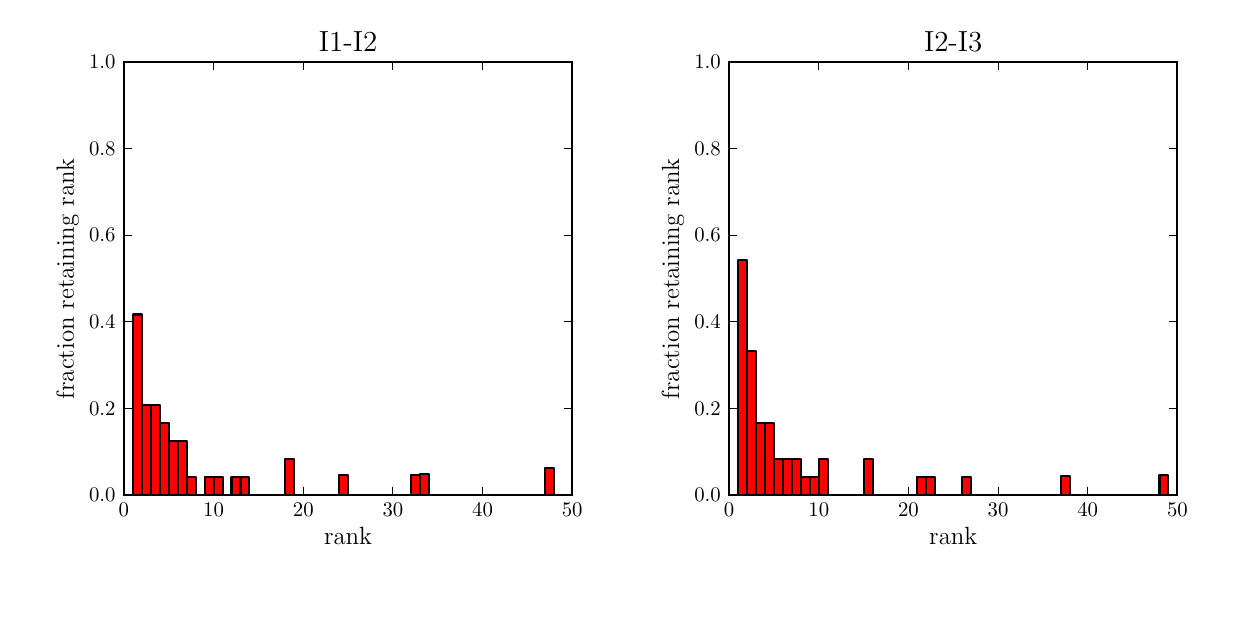}
\caption{Fraction of alters who retain their rank, as a function of rank. Left panel displays alters who retain their ranks from $I_1$ to $I_2$, and the right panel alters retain their ranks from $I_2$ to $I_3$.
}
\label{Fig_S6}
\end{center}
\end{figure}

\begin{figure}
\begin{center}
\includegraphics[width=0.9\linewidth]{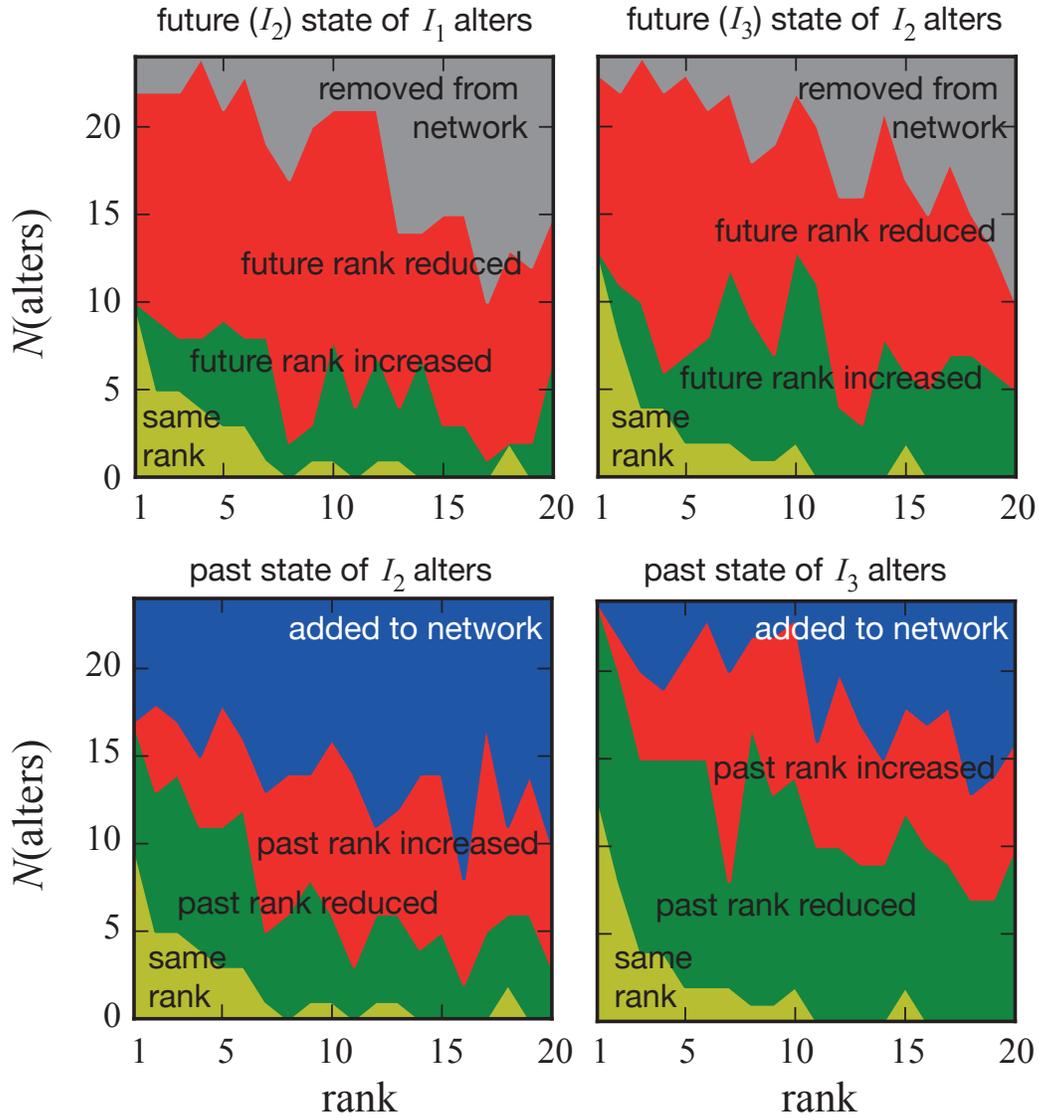}
\caption{Illustration of the changes in the states of alters of all 24 egos, from intervals $I_1$ to $I_2$ and $I_2$ to $I_3$. The top row displays the numbers of alters at a given rank in $I_1$ ($I_2$) who subsequently left the network, went down or up in ranks, or retained their rank in $I_2$ ($I_3$). The bottom row displays the number of alters at each rank in $I_2$ ($I_3$), who are new to the network, went up or down in ranks or retained their rank.
}
\label{Fig_S7}
\end{center}
\end{figure}

\end{document}